\documentclass[aps,prc,reprint,superscriptaddress]{revtex4-2}

\usepackage{graphicx}
\usepackage{dcolumn}
\usepackage{bm}
\usepackage{longtable}
\usepackage{amsmath,amssymb}
\usepackage{mathrsfs}
\usepackage{flushend}
\usepackage{microtype}
\usepackage{hyperref}
\hypersetup{
    colorlinks=true,
    linkcolor=blue,
    filecolor=blue,      
    urlcolor=blue,
    citecolor=red,
}

\begin{document}

\title{Covariant density functional theory for nuclear fission based on two-center harmonic oscillator basis}

\author{Zeyu Li}
\affiliation{China Nuclear Data Center, China Institute of Atomic Energy, Beijing 102413, China}
\affiliation{School of Physical Science and Technology, Southwest University, Chongqing 400715, China}

\author{Shengyuan Chen}
\affiliation{School of Physical Science and Technology, Southwest University, Chongqing 400715, China}

\author{Minghui Zhou}
\affiliation{School of Physical Science and Technology, Southwest University, Chongqing 400715, China}

\author{Yongjing Chen}
\email{ahchenyj@126.com}
\affiliation{China Nuclear Data Center, China Institute of Atomic Energy, Beijing 102413, China}

\author{Zhipan Li}
\email{zpliphy@swu.edu.cn}
\affiliation{School of Physical Science and Technology, Southwest University, Chongqing 400715, China}

\date{\today}

\begin{abstract}
  \begin{description}
    \item[Background] Nowdays, modern microscopic approaches for fission are generally based on the framework of nuclear density functional theory (DFT), which has enabled a self-consistent treatment of both static and dynamic aspects of fission. The key issue is a DFT solver with high precision and efficiency especially for the large elongated configurations.

    \item[Purpose] To develope a DFT solver with high precision and efficiency based on the point coupling covariant density functional theory (CDFT), which has achieved great success in describing properties of nuclei for the whole nuclear chart.

    \item[Method] We have extended the point-coupling CDFT to be based on the two-center harmonic oscillator (TCHO) basis, which matches well with the large elongated configurations during the fission process. Multi-dimensional constraint and time-dependent generator coordinate method (TDGCM) have been used to analyze the fission potential energy surface and fission dynamics, respectively. To simulate the splitting process of the nascent fragments beyond scission, we also introduce a density constraint into the new CDFT framework.

    \item[Results] Illustrative calculations have been done for the PESs and induced fission dynamics of two typical examples: $^{226}$Th and $^{240}$Pu. A more reasonable PES is obtained in the new framework compared to that based on the once-center harmonic oscillator (OCHO) with the same basis space. An optimization of about $0.2\sim0.3$ MeV has been achieved for the outer fission barriers and large elongated configurations. The dynamical simulations based on TCHO basis presents a trend to improve the description for fission yields.

    \item[Conclusions] The new developed CDFT solver optimizes the elongated configurations, improves the calculation efficiency, and provides a basis for large-scale multi-dimensional constraint calculations and dynamical simulations.

    \end{description}
  \end{abstract}

\maketitle

\section{Introduction}\label{Introduction}

Nuclear fission presents a unique example of non-equilibrium large-amplitude collective motion where all nucleons participate with complex correlation effects, making the microscopic description of fission one of the most complex problems in low-energy theoretical nuclear physics \cite{Schunck2016RPP,Hans2012Book}. Since the discovery of nuclear fission, various theories have been put forward and made great progress. Based on the work of Bohr and Wheeler \cite{Bohr1939}, the early theories for fission introduced a set of deformation parameters into the liquid drop model to construct multi-dimensional potential energy surfaces (PESs) to describe the relationship between nuclear deformation and energy,  which gives a simple explanation of nuclear fission.
In the subsequent studies, the shell corrections and pair correlations have been added to the liquid drop model, which is called the macroscopic-microscopic (MM) approach \cite{BRACK1972RMP,Nix1972ARNS}. The MM approach has a series of versions characterized by different parametrizations of nuclear surface of the liquid drop and different phenomenological nuclear potentials, such as the five-dimensional finite-range liquid-drop model (FRLDM) \cite{Moller2001Nature,Moller2009PRC,Ichikawa2012PRC}, macroscopic-microscopic Woods-Saxon model \cite{Jachimowicz2012PRC,Jachimowicz2013PRC}, the macroscopic–microscopic Lublin–Strasbourg drop (LSD) model in the three-quadratic-surface parametrization \cite{Wang2019CTP,Zhu2020CTP}, the LSD in Fourier shape parametrization \cite{Schmitt2017PRC}, two-center shell model \cite{Liu2019PRC} and so on. Based on a large number of parameters, the MM approach has greatly optimized the description of atomic nuclei. However, due to the parameter dependence of the results, the explanation of the microscopic mechanism of fission is still eludes us.

Nowdays, modern microscopic approaches for fission are generally based on the framework of nuclear density functional theory (DFT), which has enabled a self-consistent treatment of both static and dynamic aspects of fission \cite{Schunck2016RPP,Schmidt2018RPP,Simenel2018PPNP,Bender2020JPG,Verriere2020FP,Schunck2022PPNP}. In the DFT framework that relys on the adiabatic approximation, the total energies and wave functions along the fission path are generally determined by the minimization of the energy density functional of the nucleus within a given set of constraints and assumed symmetries. Then the fission observables can be obtained by performing a time-dependent evolution of the collective wave packet on the microscopic PES using, e.g. the time-dependent generator coordinate method \cite{Berger1984NPA,Goutte2005PRC,Regnier2016PRC,Regnier2016CPC,Zdeb2017PRC,Regnier2019PRC,Younes2012LLNL,TaoH2017PRC,ZhaoJ2019PRC,ZhaoJ2019PRC2,ZhaoJ2021PRC,ZhaoJ2022PRC,chen_energy_2022,chen_microscopic_2023,Chen_pair_2023,Schunck_oddA_2023}. In such approach the dynamics of the fissioning system essentially depends on the microscopic inputs, e.g. the PES and  collective inertia as functions of few collective coordinates. However, the fully microscopic and nonadiabatic time-dependent DFT has shown that many collective degrees of freedom are excited in the fission process \cite{Bulgac2016PRL}. Therefore, to achieve a better description of fission dynamics based on DFT, one needs to carry out a larger-scale multi-dimensional calculation including more collective degrees of freedom. To this end, a DFT solver with high precision and efficiency especially for the large elongated configurations is necessary.

At present, based on the nonrelativistic density functionals, the popular used DFT solvers for nuclear fission include, e.g. the codes \emph{HFBTHO} \cite{Stoitsov2013CPC}  and \emph{HFODD} \cite{Schunck2012CPC,Schunck2017CPC} that solve the Skyrme Hartree–Fock–Bogolyubov (HFB) equations in the Cartesian deformed harmonic-oscillator (HO) basis, the code \emph{SkyAx} \cite{Reinhard2021CPC} solving the Skyrme-Hartree-Fock equations on a two-dimensional mesh assuming axial symmetry, the solvers for HFB equation with the finite-range Gogny effective interaction in the deformed HO basis \cite{Warda2002PRC,Younes2009PRC}. Based on the relativistic (covariant) framework, the multi-dimensionally  constrained covariant DFT (CDFT) in an axially deformed HO basis has been implemented \cite{Lu2014PRC,TaoH2017PRC}, and very recently, the time-dependent CDFT in three-dimensional lattice space was also developed by means of the inverse Hamiltonian and spectral methods \cite{Ren2020PRC,Ren2022PRL}. The constrained CDFT in deformed HO basis has been extensively used to study the spontaneous and induced fission dynamics, and achieved acceptable agreement with the experimental data \cite{ZhaoJ2015PRC,ZhaoJ2016PRC,TaoH2017PRC,ZhaoJ2019PRC,ZhaoJ2019PRC2,ZhaoJ2021PRC,ZhaoJ2022PRC,LiZY2022PRC}. However, due to the mismatch between the one-center HO basis and the elongated configurations of the fissioning nucleus, both the accuracy and efficiency of the calculations decrease with the increase of elongation, and in particular, the calculation becomes unreliable for the configurations beyond scission. One way to solve this problem is to extend the present framework to be based on the two-center HO (TCHO) basis \cite{Berger1985PhD,Geng2007CPL,Geng2007PhD}, which has been proved to be particularly well suited for the description of highly elongated systems \cite{Berger1985PhD,Regnier2016PRC,Regnier2019PRC}. 

In this work, we will extend the point-coupling CDFT to be based on the TCHO basis and perform illustrative calculations for the PESs and induced fission dynamics of two typical examples: $^{226}$Th and $^{240}$Pu. Moreover, we will implement the density constraint in the new developed CDFT and analyze the configurations and potential energy curve beyond scission. In Sec. \ref{Theory}, the theoretical framework is introduced. The results for PESs and fragment yield distributions calculated based on CDFT in one-center and two-center HO basis are compared and discussed in detail in Sec. \ref{Benchmark}. Density-constrained calculation for post-scission configurations is briefly discussed in Sec. \ref{Density constrained DFT}. Sec. \ref{Summary} contains a summary of results and an outlook for future studies.

\section{Theoretical framework}\label{Theory}

\subsection{Covariant density functional theory}

The energy density functional in the point-coupling version for the CDFT can be written as
\begin{gather}
  \begin{split}
  {{E}_{\rm CDF}}&=\int{{d}}\mathbf{r}{{\varepsilon }_{\rm CDF}}(\mathbf{r}) \\ 
             &=\sum\limits_{k}{\int{d}}\mathbf{r}\upsilon _{k}^{2}{{{\bar{\psi }}}_{k}}(\mathbf{r})(-i\mathbf{\gamma }\mathbf{\nabla} +m){{\psi }_{k}}(\mathbf{r}) \\ 
             &+\int{d\mathbf{r}\left( \frac{{{\alpha }_{S}}}{2}\rho _{S}^{2}+\frac{{{\beta }_{S}}}{3}{{\rho}_S^{3}}+\frac{{{\gamma }_{S}}}{4}\rho _{S}^{4}+\frac{{{\delta }_{S}}}{2}{{\rho }_{S}}\Delta {{\rho }_{S}} \right.} \\ 
             &+\frac{{{\alpha }_{V}}}{2}{{j}_{\mu }}{{j}^{\mu }}+\frac{{{\gamma }_{V}}}{4}{{({{j}_{\mu }}{{j}^{\mu }})}^{2}}+\frac{\delta_V}{2}{{j}_{\mu }}\Delta {{j}^{\mu }}+\frac{{e}}{2}{{\rho }_{p}}{{A}^{0}} \\ 
             &\left. +\frac{{{\alpha }_{TV}}}{2}j_{TV}^{\mu }\cdot {{(j_{TV})}_{\mu }}+\frac{{{\delta }_{TV}}}{2}j_{TV}^{\mu }\cdot \Delta (j_{TV})_{\mu } \right) \label{EDF}
  \end{split}  
\end{gather}
with the local densities and currents
\begin{gather}
  \begin{split}
  {{\rho }_{S}}(\mathbf{r})&=\sum\limits_{k}{v_{k}^{2}}{{\bar{\psi }}_{k}}(\mathbf{r}){{\psi }_{k}}(\mathbf{r})\\
  {{j}^{\mu }}(\mathbf{r})&=\sum\limits_{k}{v_{k}^{2}}{{\bar{\psi }}_{k}}(\mathbf{r}){{\gamma }^{\mu }}{{\psi }_{k}}(\mathbf{r})\\
  j_{TV}^{\mu }(\mathbf{r})&=\sum\limits_{k}{v_{k}^{2}}{{\bar{\psi }}_{k}}(\mathbf{r}){{\gamma }^{\mu }}\tau_3{{\psi }_{k}}(\mathbf{r})
\end{split}
\label{eq:density}
\end{gather}
where $\psi$ is the Dirac spinor field of the nucleon. $\rho_p$ and $A^0$ are respectively the proton density and Coulomb field. The subscripts indicate the symmetry of the couplings: $S$ stands for scalar, $V$ for vector, and $T$ for isovector. Various coupling constants $(\alpha,\beta,\gamma,\delta)$ are determined by the PC-PK1 parametrization.\cite{Zhao2010PRC}. 

Minimizing the energy density functional Eq. (\ref{EDF}) with respect to $\bar{\psi}_k$, one obtains the Dirac equation for the single nucleons
\begin{gather}
  \left\{-i\mathbf{\alpha}\cdot\mathbf{\nabla} +V(\mathbf{r})+\beta [M+S(\mathbf{r})]\right\}\psi_k(\mathbf{r})=\varepsilon_k\psi_k(\mathbf{r}) \label{Dirac equ1}
\end{gather}
where the local scalar $S(r)$ and vector $V(r)$ potentials are functions of densities and currents in the nucleus
\begin{gather}
  \begin{split}
  S(\mathbf{r})&=\alpha_S \rho_S+\beta_S\rho_S^2+\gamma_S\rho_S^3+\delta_S\Delta\rho_S\\
  V^\mu (\mathbf{r})&=\alpha_V j^\mu+\gamma_V(j_\nu j^\nu)j^\mu+\delta_V\Delta j^\mu+eA^\mu\frac{1-\tau_3}{2}\\
           & \ \ \ +\tau_3(\alpha_{TV}j^\mu_{TV}+\delta_{TV}\Delta j_{TV}^\mu)
  \end{split}
  \label{eq:SV}
\end{gather}
Solving the Eqs. (\ref{eq:density}-\ref{eq:SV}) iteratively, one can obtain the single-nucleon wave functions, densities and currents, and also the binding energy of the nucleus $E_{\rm CDF}$. Here, it is noted that the Broyden method \cite{Broyden1965MC,Baran2008PRC} has been used in the iteration, which can speed up the convergency by about one order compared the linear mixing for the elongated configurations.

Pairing correlations between nucleons are treated using the Bardeen-Cooper-Schrieffer (BCS) approach with a $\delta$ pairing force \cite{Burvenich2002PRC}. Due to the broken of the translational symmetry, one has to consider the center-of-mass (c.m.) correction energy for the motion of the c.m. and here a phenomenological formulas $E_{\rm c.m.}=-\frac{3}{4}\cdot 41A^{-1/3}$ is adopted. Finally, the total energy reads
\begin{gather}
  E_{\rm tot}=E_{\rm CDF}+E_{\rm pair}+E_{\rm c.m.}
\end{gather}

\subsection{\label{subsec:TCHO}Two-center harmonic oscillator basis}

To calculate the multi-dimensional PES in a large deformation space, one needs to solve the Dirac equation (\ref{Dirac equ1}) with high precision and efficiency. Here we will expand the Dirac spinor in a two-center harmonic oscillator (TCHO) basis to match the large elongated configuration during fission. The axially symmetric TCHO potential in cylindrical coordinate system reads
\begin{gather}
  V\left(r_{\perp}, z\right)=\frac{1}{2} M \omega_{\perp}^{2} r_{\perp}^{2}+ \begin{cases}\frac{1}{2} M \omega_{1}^{2}\left(z+z_{1}\right)^{2}, & z<0 \\ \frac{1}{2} M \omega_{2}^{2}\left(z-z_{2}\right)^{2}, & z \geq 0\end{cases}
\end{gather}
where $M$ is the nucleon mass. TCHO can be regarded as two off-center harmonic oscillators connected at $z=0$, while $z_1 (z_2)$ and $\omega_1 (\omega_2)$ denote the distance from $z=0$ to the center of the left (right) harmonic oscillator and its frequency, respectively.

Due to the spatial rotational symmetry along $z$ axis, the eigenfunction of TCHO can be written as the product of eigenfunctions of different degrees of freedom
\begin{gather}
  \Phi \left({{r}_{\bot }}, z,\varphi, s  \right)=\phi _{{{n}_{r}}}^{{{m}_{l}}}\left( {{r}_{\bot }} \right){{\phi }_{\nu }}(z)\frac{1}{\sqrt{2 \pi}} e^{ i m_{l} \varphi} \chi _{m_s}
\end{gather}
with
\begin{gather}
  \phi _{{{n}_{r}}}^{{{m}_{l}}}\left( {{r}_{\bot }} \right)=\frac{\sqrt{2}}{{b}_{\bot }}\sqrt{\frac{n_r!}{(n_r+m_l)!}}{{\eta }^{{{m}_{l}}/2}}L_{{{n}_{r}}}^{{{m}_{l}}}(\eta ){{e}^{-\eta /2}}
\end{gather}
\begin{gather}
  \phi_{\nu}(z)=\left\{\begin{array}{l}
    C_{\nu_{1}} H_{\nu_{1}}\left(-\zeta_{1}\right) e^{-\zeta_{1}^{2} / 2} \ \ \ \text {for }    z<0 \\ 
    C_{\nu_{2}} H_{\nu_{2}}\left(\zeta_{2}\right) e^{-\zeta_{2}^{2} / 2} \ \ \ \  \text { for }    z \geq 0\end{array}\right.
  \label{eq:phiz}
\end{gather}
where $\eta =r_{\bot}^2/{b_{\bot}^2}$, $\zeta_{1}=\left(z+z_{1}\right) / b_{1}$, and $\zeta_{2}=\left(z-z_{2}\right) / b_{2}$. $b_{\bot}$, $b_1$ and $b_2$ are the characteristic lengths obeying the general relationship $b=\sqrt{\hbar/M\omega}$ with their corresponding frequencies. $L_{n_r}^{m_l}(\eta)$ and $H_{\nu}(\zeta)$  denote the associated Laguerre polynomial and Hermite function, respectively. In Eq. (\ref{eq:phiz}), $\nu_1$, $\nu_2$, $C_{\nu_{1}}$ and $C_{\nu_{2}}$ are determined by four conditions: continuity of $\phi_{\nu}(z)$ and $\phi^\prime_{\nu}(z)$ at $z=0$, stationary condition of eigen energy, and normalization \cite{Geng2007PhD}. For the convenience of discussion and application, we set $z_1=z_2$, $b_1=b_2$ and denote the TCHO basis as $|\alpha\rangle=|n_r\nu m_l m_s\rangle$ in the following. The choice of parameters $z_1$ and $b_1$ has been introduced in detail in Appendix \ref{appendixA}.

\begin{figure}[htbp]
  \centering
  \includegraphics[width=8cm]{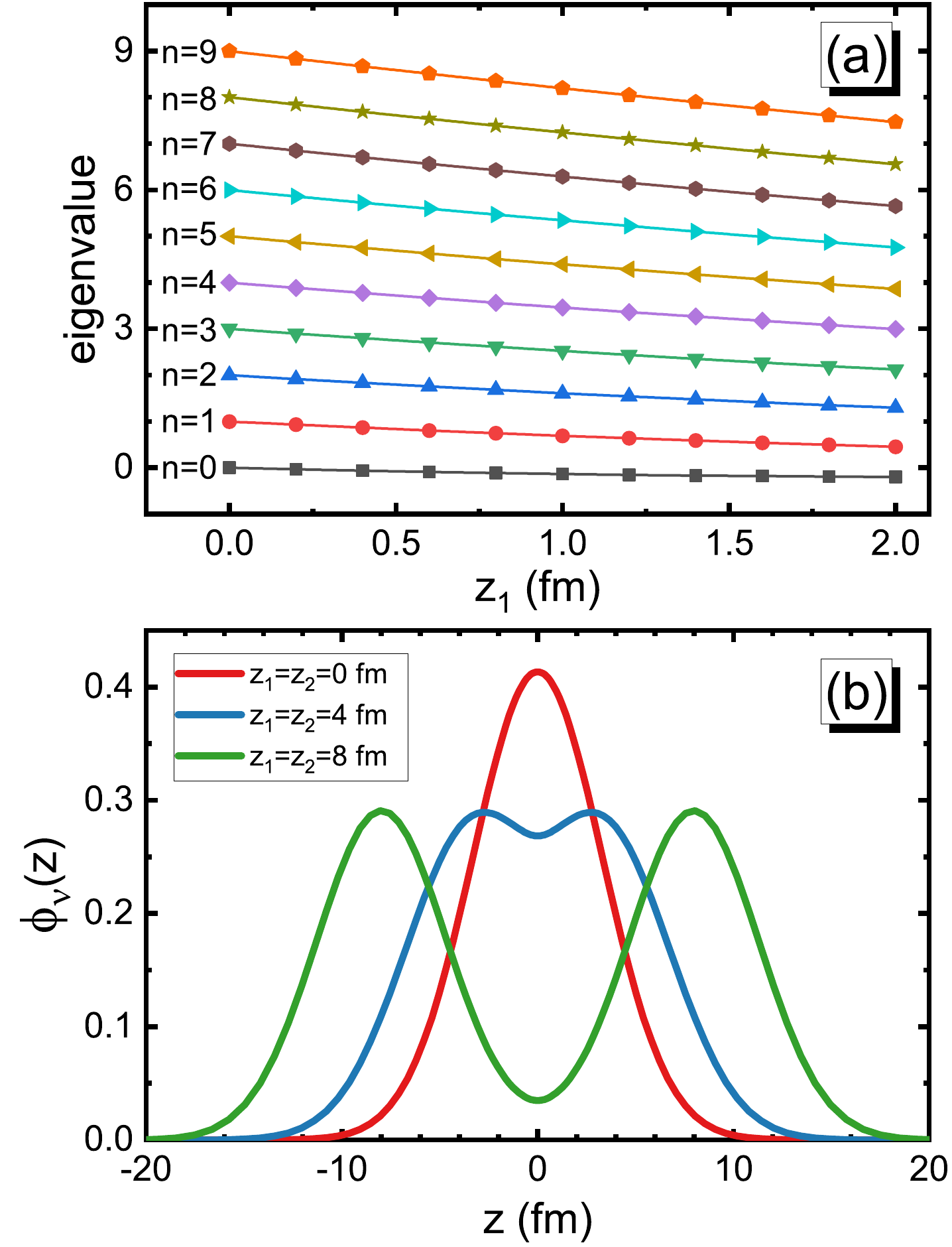}
  \caption{ (Color online) Evolution of the eigenvalue $\nu_1$ (a) and the wave function $\phi_\nu(z)$ of the ground state (b) as increasing $z_1$ for fixed $b_1=3.3$ fm.}
  \label{eigen}
\end{figure}

To get an intuitional impression of the TCHO basis, in Fig. \ref{eigen}, we show the evolution of the eigenvalue $\nu_1$ and the wave function $\phi_\nu(z)$ of the ground state as increasing $z_1$. Obviously, the eigen quantum numbers are not integers when $z_1\neq 0$ and the levels become denser for larger $z_1$. Remarkably, the evolution of wave function is consistent with that of the configurations in the fission precess. 

To solve the Dirac equation (\ref{Dirac equ1}) in the TCHO basis, firstly we expand the Dirac spinor $\psi_k$ as
\begin{equation}
\label{eq:psi}
\psi_k(r,s)=\left(\begin{array}{c} f_k(r,s) \\ ig_k(r,s) \end{array} \right)
                =\left(\begin{array}{c} \sum\limits_\alpha f^k_\alpha |\alpha\rangle \\
                i \sum\limits_{\bar\alpha} g^k_{\bar\alpha}|\bar\alpha\rangle\end{array} \right) \; .
\end{equation}
The summation of $\alpha$ has to be truncated for a given number of shells $N_f$ which satisfies $E_\alpha\leq (N_f+3/2) \hbar\omega$ with $\hbar\omega=41A^{-1/3}$ MeV, and the summation of $\bar\alpha$ is truncated at $N_g=N_f+1$ to avoid spurious states. Then we obtain the Dirac equation in matrix form
\begin{equation}
  \left(\begin{array}{cc}
  A_{\alpha, \alpha^{\prime}} & B_{\alpha, \bar{\alpha}^{\prime}} \\
  B_{\bar{\alpha}, \alpha^{\prime}} & -C_{\bar{\alpha}, \bar{\alpha}^{\prime}}
  \end{array}\right)\left(\begin{array}{c}
  f_{\alpha^{\prime}}^k \\
  g_{\bar{\alpha}^{\prime}}^k
  \end{array}\right)=\varepsilon _k\left(\begin{array}{c}
  f_\alpha^k \\
  g_{\bar{\alpha}}^k
  \end{array}\right)
  \label{Dirac_equ_matrix}
\end{equation}
The matrix elements $A_{\alpha, \alpha^{\prime}}$, $B_{\alpha, \bar{\alpha}^{\prime}}$ and $C_{\bar{\alpha}, \bar{\alpha}^{\prime}}$ can be expressed as follows
\begin{equation}
  \begin{aligned}
  A_{\alpha, \alpha^{\prime}} = & \delta_{m_l m_l^{\prime}} \delta_{m_s m_s^{\prime}} N_{n_r}^{m_l} N_{n_{r}^\prime}^{m_l} \int_0^{\infty} d \eta e^{-\eta} \eta^{m_l} L_{n_r}^{m_l}(\eta) L_{n_r^\prime}^{m_l}(\eta) \\
  & \times \left[C_{\nu_1} C_{\nu_{1}^{\prime}} \int_{-\infty}^0 d z e^{-\xi_1^2} H_{\nu_1}\left(-\xi_1\right) H_{\nu_1^\prime}\left(-\xi_1\right)\right.\\
  & \left.\quad+C_{\nu_2} C_{\nu_2^{\prime}} \int_0^{\infty} d z e^{-\xi_2^2} H_{\nu_2}\left(\xi_2\right) H_{\nu_2^\prime}\left(\xi_2\right)\right]\\
  & \times\left(M+S(r_{\perp}, z)+V(r_{\perp}, z)\right)
  \end{aligned}
\end{equation}

\begin{equation}
  \begin{aligned}
    C_{\bar{\alpha}, \bar{\alpha}^{\prime}} = & \delta_{m_l m_l^{\prime}} \delta_{m_s m_s^{\prime}} N_{n_r}^{m_l} N_{n_{r}^\prime}^{m_l} \int_0^{\infty} d \eta e^{-\eta} \eta^{m_l} L_{n_r}^{m_l}(\eta) L_{n_r^\prime}^{m_l}(\eta) \\
  & \times \left[C_{\nu_1} C_{\nu_{1}^{\prime}} \int_{-\infty}^0 d z e^{-\xi_1^2} H_{\nu_1}\left(-\xi_1\right) H_{\nu_1^\prime}\left(-\xi_1\right)\right.\\
  & \left.\quad+C_{\nu_2} C_{\nu_2^{\prime}} \int_0^{\infty} d z e^{-\xi_2^2} H_{\nu_2}\left(\xi_2\right) H_{\nu_2^\prime}\left(\xi_2\right)\right]\\
  & \times\left(M+S(r_{\perp}, z)-V(r_{\perp}, z)\right)
  \end{aligned}
\end{equation}

\begin{equation}
  \begin{aligned}
  B_{\alpha, \bar{\alpha}^{\prime}}= & \delta_{m_l m_l^{\prime}} \delta_{m_s m_s^{\prime}} \delta_{n_r n_r^{\prime}}(-1)^{\frac{1}{2}-m_s} I_1\left(\nu, \nu^{\prime}\right) \\
  &  +\delta_{\nu\nu^{\prime}}\delta_{m_l m_l^{\prime}-1} \delta_{m_s m_s^{\prime}+1} \frac{N_{n_r}^{m_l} N_{n_r^{\prime}}^{m_l^{\prime}}}{b_{\perp}} \int_0^{\infty} d \eta e^{-\eta}  \\
  &~~\eta^{m_l / 2}\eta^{\left(m_l^{\prime}-1\right) / 2} L_{n_r}^{m_l}(\eta)\left(\tilde{L}_{n_r^{\prime}}^{m_{l}^{\prime}}(\eta)+m_l^{\prime} L_{n_r^{\prime}}^{m_l^{\prime}}(\eta)\right)  \\
  & +\delta_{\nu\nu^{\prime}}\delta_{m_l m_l^{\prime}+1} \delta_{m_s m_s^{\prime}-1} \frac{N_{n_r}^{m_l} N_{n_r^{\prime}}^{m_l^{\prime}}}{b_{\perp}} \int_0^{\infty} d \eta e^{-\eta}   \\ 
  &~~\eta^{m_l / 2}\eta^{\left(m_l^{\prime}-1\right) / 2} L_{n_r}^{m_l}(\eta)\left(\tilde{L}_{n_r^{\prime}}^{m_{l}^{\prime}}(\eta)-m_l^{\prime} L_{n_r^{\prime}}^{m_l^{\prime}}(\eta)\right)  \\
  \end{aligned}
\end{equation}
with
\begin{equation}
  \begin{aligned}
  I_1\left(\nu, \nu^{\prime}\right)= &  \frac{C_{\nu_1}C_{\nu_1^{\prime}}}{b_1} \int_{-\infty}^0 d z e^{-\xi_1^2} H_{\nu_1}\left(-\xi_1\right)\\
  &\left(\xi_1 H_{\nu_1}\left(-\xi_1\right)+H_{\nu_1^{\prime}+1}\left(-\xi_1\right)\right) \\
  & + \frac{C_{\nu_2}C_{\nu_2^\prime}}{b_2} \int_0^{\infty} d z e^{-\xi_2^2} H_{\nu_2}\left(\xi_2\right)\\
  &\left(\xi_2 H_{\nu_2}\left(\xi_2\right)-H_{\nu_2^{\prime}+1}\left(\xi_2\right)\right)
  \end{aligned}
  \end{equation}
Finally, we can obtain the single nucleon energies and wave functions by diagonalizing the Hamiltonian matrix. In the following we will denote the new implementation of CDFT based on TCHO basis as CDFT-TCHO.

\subsection{Multi-dimensional constraint calculation}

The entire map of the energy surface in multi-dimensional collective space for fission is obtained by imposing constraints on a number of collective coordinates, e.g. axial quadrupole and octupole moments $q_2$, $q_3$, and the number of nucleons in the neck $q_N$
\begin{equation}
  \langle E_{\rm tot}\rangle+\sum\limits_{k=2,3}C_k(\langle\hat Q_k\rangle-q_k)^2+C_N(\langle\hat Q_N\rangle-q_N)^2,
\end{equation}
where $\langle E_{\rm tot}\rangle$ is the total energy of CDFT, $C_k$ and $C_N$ are the corresponding stiffness constants. $\hat Q_2$, $\hat Q_3$, and $\hat Q_N$ denote the mass quadrupole and octupole operators, and the Gaussian neck operator, respectively
\begin{gather}
  \begin{aligned}
  \hat{Q}_2&=2z^2-r_\perp^2\\
  \hat{Q}_3&=2z^3-3zr_\perp^2\\
  \hat Q_N &=e^{-(z-z_N)^2/a^2_N}
\end{aligned}
\end{gather}
where $a_N$ = 1 fm and $z_N$ is the position of the neck determined by minimizing $\langle\hat Q_N\rangle$ \cite{Younes2009PRC}. The left and right fragments are defined as parts of the whole nucleus with $z\leq z_N$ and $z\geq z_N$, respectively.

The widely used quadrupole and octupole deformation parameters $\beta_2$ and $\beta_3$ can be determined from the following relations:
\begin{gather}
  \beta_2=\frac{\sqrt{5\pi}}{3AR_0^2}\langle \hat{Q}_2\rangle\\
  \beta_3=\frac{\sqrt{7\pi}}{3AR_0^3}\langle \hat{Q}_3\rangle
\end{gather}
with $R_0=r_0A^{1/3}$ and $r_0=1.2$ fm.

When the configuations for the full collective space are obtained under the constraint calculations, we can finally determine the collectve PES by subtracting the energy of zero-point motion, e.g. the vibrational and rotational zero-point motions
\begin{equation}
  \label{eq:Vcoll}
  V(\beta_2, \beta_3, q_N, \cdots)=E_{\rm tot}-\Delta E_{\rm vib}-\Delta E_{\rm rot}.
  \end{equation}
The zero-point energy (ZPE) corrections are calculated in the cranking approximation \cite{Girod1979NPA} and the expression for vibrational ZPE reads
\begin{equation}
  \label{ZPE-vib}
  \Delta E_{\rm vib} = \frac{1}{4} \textnormal{Tr}\left[\mathcal{M}_{(3)}^{-1}\mathcal{M}_{(2)}  \right]\;,
  \end{equation}
with
\begin{equation}
  \label{masspar-M}
  \mathcal{M}_{(n),kl}=\sum_{i,j}
   {\frac{\left\langle i\right|\hat{Q}_{k}\left| j\right\rangle
   \left\langle j\right|\hat{Q}_{l}\left| i\right\rangle}
   {(E_i+E_j)^n}\left(u_i v_j+ v_i u_j \right)^2}\;.
  \end{equation}
where $E_i$ and $v_i$ are the quasiparticle energies and occupation probabilities, respectively. The summation is over the proton and neutron single-particle states in the canonical basis. The rotational ZPE takes the form
\begin{equation}
  \label{ZPE-rot}
  \Delta E_{\rm rot}=\frac{\langle\hat J^2\rangle}{2{\cal I}}\;,
  \end{equation}
where ${\cal I}$ is the Inglis-Belyaev moment of inertia \cite{Inglis1956PR,Beliaev1961NP}.

For the dynamical simulation, one also needs to calculate the mass tensor in the perturbative cranking approximation \cite{Girod1979NPA} 
\begin{equation}
\label{eq:BB}
B_{kl}(\beta_2, \beta_3, q_N, \cdots)=\hbar^2 \left[\mathcal{M}_{(1)}^{-1} \mathcal{M}_{(3)} \mathcal{M}_{(1)}^{-1}\right]_{kl}.
\end{equation}

\subsection{Time-dependent generator coordinate method}

Based on the adiabatic approximation for low-energy fission, the nucleon degree of freedom is decoupled from the  collective degrees of freedom, and therefore, we can adopt the time-dependent generator coordinate method (TDGCM) with a few collective coordinates to describe the fission dynamics \cite{Berger1984NPA}. In the TDGCM approach, the nuclear wave function is described as a linear superposition of many-body functions parametrized by a vector of collective coordinates, e.g. quadrupole and octupole deformations $\beta_2$, $\beta_3$. When employing the Gaussian overlap approximation (GOA), the GCM Hill-Wheeler equation reduces to a local,  time-dependent Sch{\" o}dinger-like equation in collective space
\begin{align}
  & i \hbar \frac{\partial}{\partial t} g(\beta_{2}, \beta_{3}, t) \notag \\
  & = \left[-\frac{\hbar^{2}}{2} \sum_{k l} \frac{\partial}{\partial \beta_{k}} B^{-1}_{k l}\left(\beta_{2}, \beta_{3}\right) \frac{\partial}{\partial \beta_{l}}+ V(\beta_{2}, \beta_{3})\right] \notag \\
  &\ \ \ \times g(\beta_{2}, \beta_{3}, t) \label{eq:Schrodinger_like}
\end{align}
where $g(\beta_2, \beta_3, t)$ is a complex wave function, which contains all the information about the dynamics of the system. $V(\beta_2, \beta_3)$ and $B_{kl}(\beta_2, \beta_3)$ are the collective potential and mass tensor, respectively, and they completely determine the dynamics of the fission process in the TDGCM+GOA framework. The probability current is defined by
\begin{align}
    J_{k}\left(\beta_{2}, \beta_{3}, t\right)= &\frac{\hbar}{2 i} \sum_{l=2}^{3} B^{-1}_{k l}\left(\beta_{2}, \beta_{3}\right)\left[g^{*}\left(\beta_{2}, \beta_{3}, t\right)\frac{\partial g\left(\beta_{2}, \beta_{3}, t\right)}{\partial \beta_{l}}\right. \notag \\
   &\left. -g\left(\beta_{2}, \beta_{3}, t\right) \frac{\partial g^{*}\left(\beta_{2}, \beta_{3}, t\right)}{\partial \beta_{l}}\right] \label{currunt_eq}
\end{align}

Starting from an initial state of the compound nucleus, the collective current will move to a large deformation region and pass through a so-called scission line that is composed of the hypersurface at which the nucleus splits. At the time $t$, the measurement of the probability of a given pair of fragments can be calculated when the flux of the probability current runs through the scission line. For a surface element $\xi$, sum of the time-integrated flux of the probability $F(\xi, t)$ can be read as \cite{Regnier2018CPC}:
  \begin{equation}
    F(\xi, t)=\int_{t=0}^{t} d t \int_{\left(\beta_{2}, \beta_{3}\right) \in \xi} \mathbf{J}\left(\beta_{2}, \beta_{3}, t\right) \cdot d \mathbf{S}  \label{eq:tot_flux}
  \end{equation}
For each point on the scission line, it contains the information of $(A_L, A_H)$ which represent the masses of light and heavy fragments, respectively. Hence the yield of fission fragments with mass $A$ can be defined formally as:
  \begin{equation}
    Y(A) \propto \sum_{\xi \in \mathcal{A}} \lim _{t \rightarrow+\infty} F(\xi, t) \label{yields_eq}
  \end{equation}
where $\mathcal{A}$ is the set of all elements $\xi$ belonging to the scission line such that the heavy  or light fragment has mass $A$. Here, we  will use the software package FELIX-2.0 \cite{Regnier2018CPC} to solve the time-dependent Sch{\" o}dinger-like equation and calculate the fission observables.

\section{Illustrative calculations for $^{226}$Th and $^{240}$Pu}\label{Benchmark}

In this section, we present the illustrative calculations for two typical examples: $^{226}$Th and $^{240}$Pu. Specifically, we will compare the PESs, scission lines, and fragment yield distributions calculated based on CDFT in OCHO and TCHO basis to demonstrate the improvement of the computing efficiency and accuracy of CDFT-TCHO. In the CDFT framework, the energy density functional PC-PK1 \cite{Zhao2010PRC} determines the effective interaction in the particle-hole channel, and a $\delta$ force is used in the particle-particle channel. The strength parameters of the $\delta$ force are $V_n (V_p)= 360 (378)$ MeV fm$^3$ and $V_n (V_p) = 338 (372.5)$ MeV fm$^3$ for $^{226}$Th and $^{240}$Pu, respectively, which are determined by reproducing the empirical pairing gaps from a five-point formula \cite{Bender2000EPJA}. 

In the first step, a large-scale deformation-constrained CDFT calculation is performed to generate the PESs, scission lines, and mass tensors in the $\beta_2$-$\beta_3$ plane.  The range of collective variables is -0.98 to 6.98 for $\beta_2$ with a step $\Delta\beta_2=0.04$, and from 0.00 to 4.24 for $\beta_3$ with a step $\Delta\beta_3=0.08$.  When describing fission in a collective space, scission is characterized by a discontinuity between the two domains of prescissioned and postscissioned configurations. Following our previous study \cite{TaoH2017PRC}, here we define the prescission domain by the nucleon number in the neck $q_N\geq 3$ and consider the frontier of this domain as the scission line.

For the induced fission dynamics, the TDGCM+GOA is performed to model the time-evolution of the fissioning nucleus with a time step $\delta t=5\times 10^{-4}$ zs.  The parameters of the additional imaginary absorption potential that takes into account the escape of the collective wave packet in the domain outside the region of calculation are: the absorption rate $r=20\times 10^{22} \text{s}^{-1}$, and the width of the absorption band $w=1.5$.

\subsection{Results for $^{226}$Th}

\begin{figure}[htbp]
  \centering
  \includegraphics[width=8.5cm]{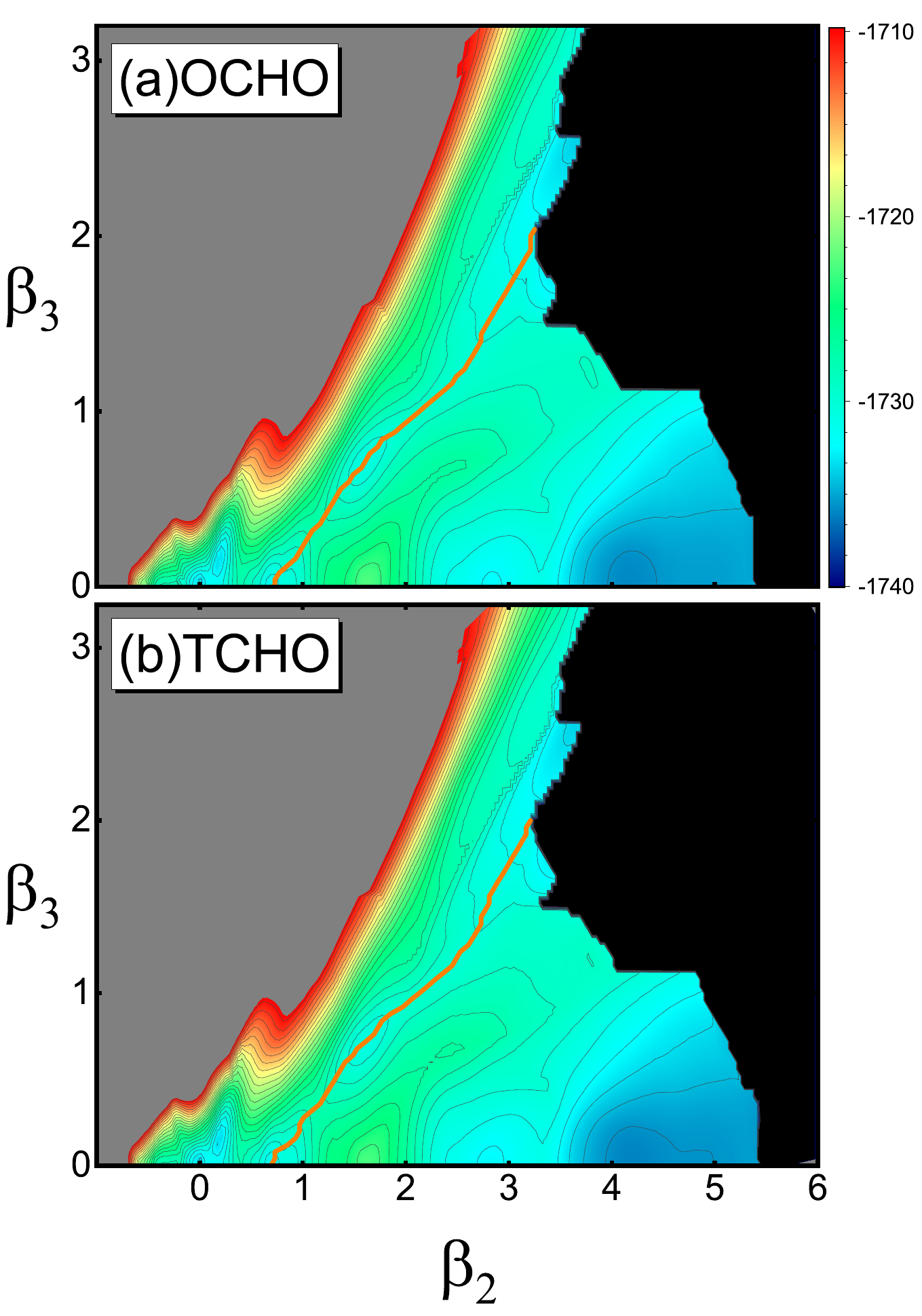}
  \caption{ (Color online) The potential energy surfaces of $^{226}\text{Th}$ in the $\beta_2$-$\beta_3$ plane calculated using the CDFT in OCHO (a) and TCHO (b) basis with a cutoff of major shell $N_f=20$. The orange solid line denotes the optimal fission path.}
  \label{PES}
\end{figure}

Fig. \ref{PES} displays the PESs of $^{226}\text{Th}$ in the $\beta_2$-$\beta_3$ plane calculated using the CDFT in OCHO and TCHO basis with a cutoff of major shell $N_f=20$. The orange solid line denotes the optimal fission path. The topography of these two PESs looks almost same on the whole, and is also comparable with that obtained using the  Hartree-Fock-Bogoliubov framework based on the Gogny D1S functional \cite{Dubray2008PRC}. Two competing fission vellays, i.e. the asymmetric one passing by the optimal fission path and the symmetric one with $\beta_3\sim 0$, are observed and they are separated by a ridge from $(\beta_2, \beta_3)\sim (1.5, 0.0)$ to $(3.8, 1.2)$.

\begin{figure}[htbp]
  \centering
  \includegraphics[width=8cm]{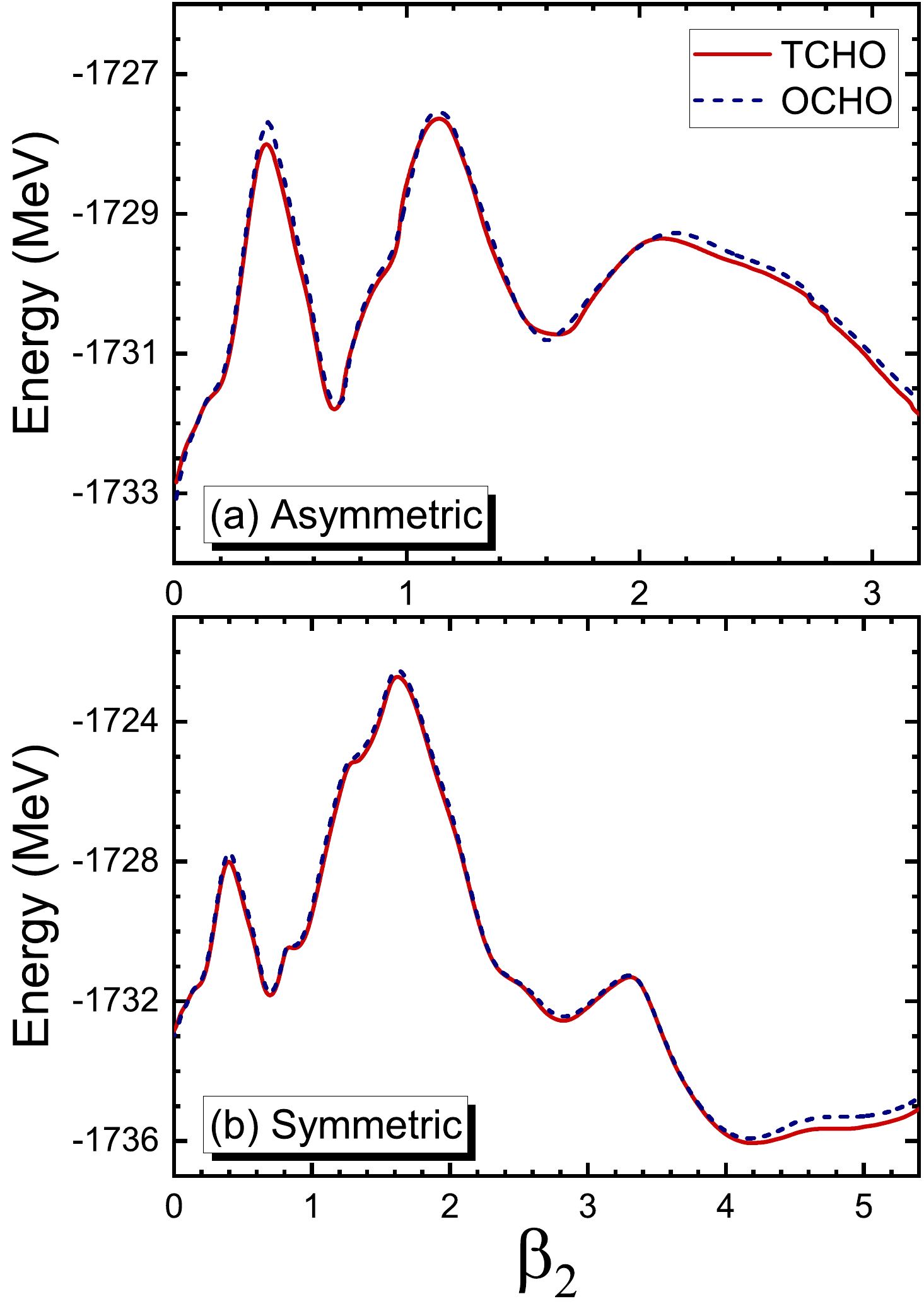}
  \caption{ (Color online) The potential energy curves along the asymmetric (a) and symmetric (b) fission paths of $^{226}$Th calculated using CDFT in TCHO and OCHO basis.}
  \label{path}
\end{figure}

To compare the two calculations in more detail, we present the potential energy curves along both the asymmetric and symmetric fission paths in Fig. \ref{path}. The potential energies for the fission barriers and large elongated configurations calculated using TCHO are generally lower than those of OCHO. Specifically, along the asymmetric path, a triple-humped fission barrier is predicted, and the calculated heights using TCHO (OCHO) are 5.92 (6.11), 6.05 (6.25), and 4.32 (4.51) MeV from the inner to the outer barrier, respectively. While, the barrier heights along the symmetric path are 5.92 (6.11), 10.99 (11.30), and 2.37 (2.52) MeV from the inner to outer. Please note that here the values for the calculation with OCHO basis are different from those in our pervious work \cite{TaoH2017PRC} due to the consideration of the vibrational and rotational ZPEs [c.f. Eq. \ref{eq:Vcoll})] in current work. Moreover, we also find that the ridge between two fission paths is lowered, e.g. $\sim 0.1$ MeV for $(\beta_2, \beta_3)= (3.46, 1.20)$.

\begin{figure}[htbp]
  \centering
  \includegraphics[width=8cm]{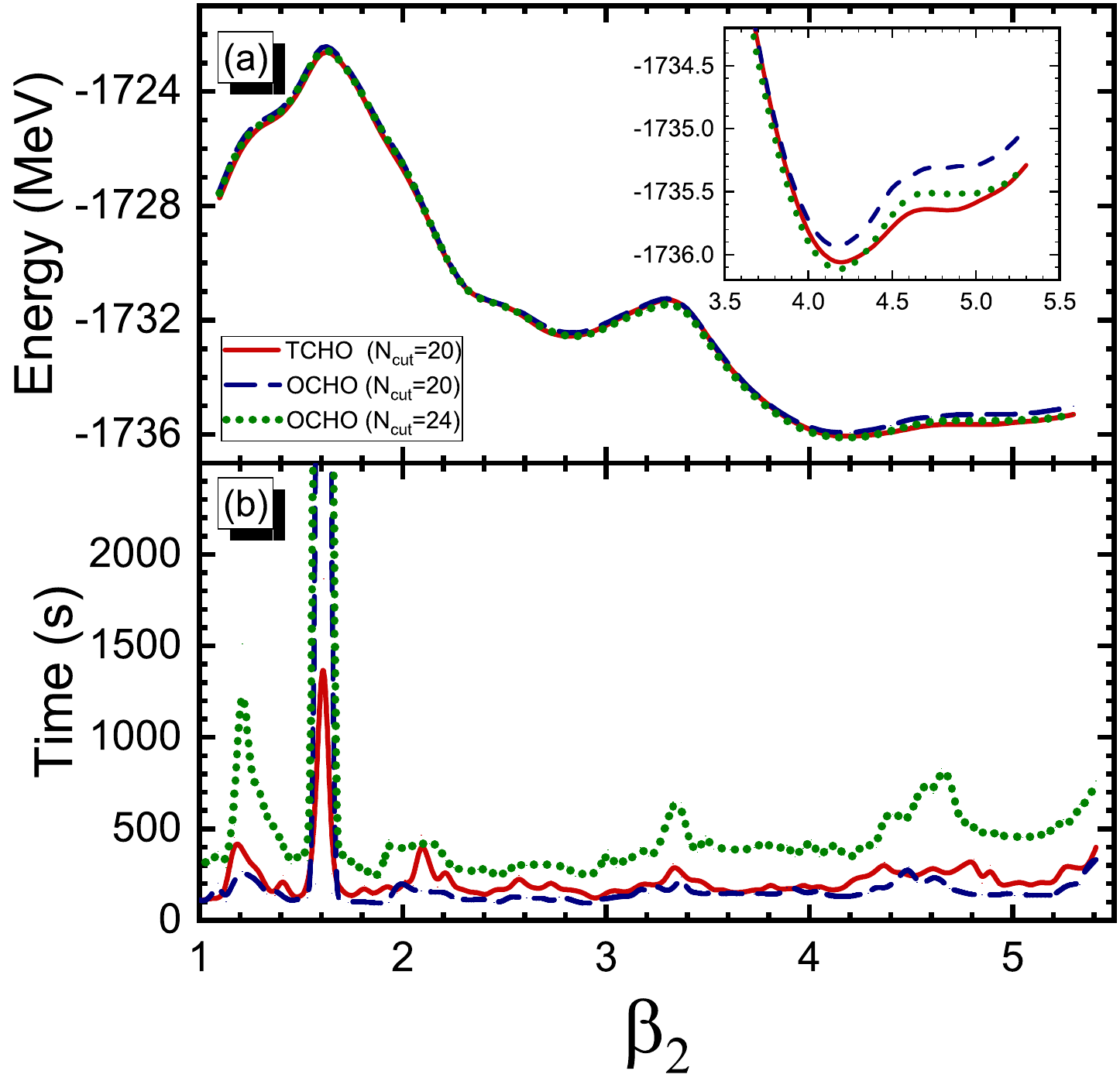}
  \caption{ (Color online) The potential energy curves (a) and computing time (b) for the symmetric fission path of $^{226}$Th calculated using CDFT in TCHO and OCHO basis. The solid, dashed, and dotted curves correspond to the results calculated from TCHO with a cutoff of major shell $N_f=20$, OCHO with $N_f=20$ and $N_f=24$, respectively. The inset of panel (a) displays the potential energy curves for large deformations.}
  \label{time}
\end{figure}

To check the efficiency of CDFT-TCHO, we take the more elongated symmetric fission path as an example to compare the computing time for three cases: TCHO with a cutoff of major shell $N_f=20$, OCHO with $N_f=20$ and $N_f=24$ in Fig. \ref{time}. For the first two calculations, the computing time is similar, $\lesssim 300$ seconds for most of the configurations, but the accuracy of TCHO is better, $\sim 0.3$ MeV lower for the large deformations due to the match between basis and calculated configurations. To get similar accuracy as that of TCHO, the OCHO basis space has to be enlarged to $N_f=24$, which of course consumes more time, about twice than that of TCHO calculation and even worse for the second fission barrier at $\beta_2\sim 1.6$.

\begin{figure}[htbp]
  \centering
  \includegraphics[width=8cm]{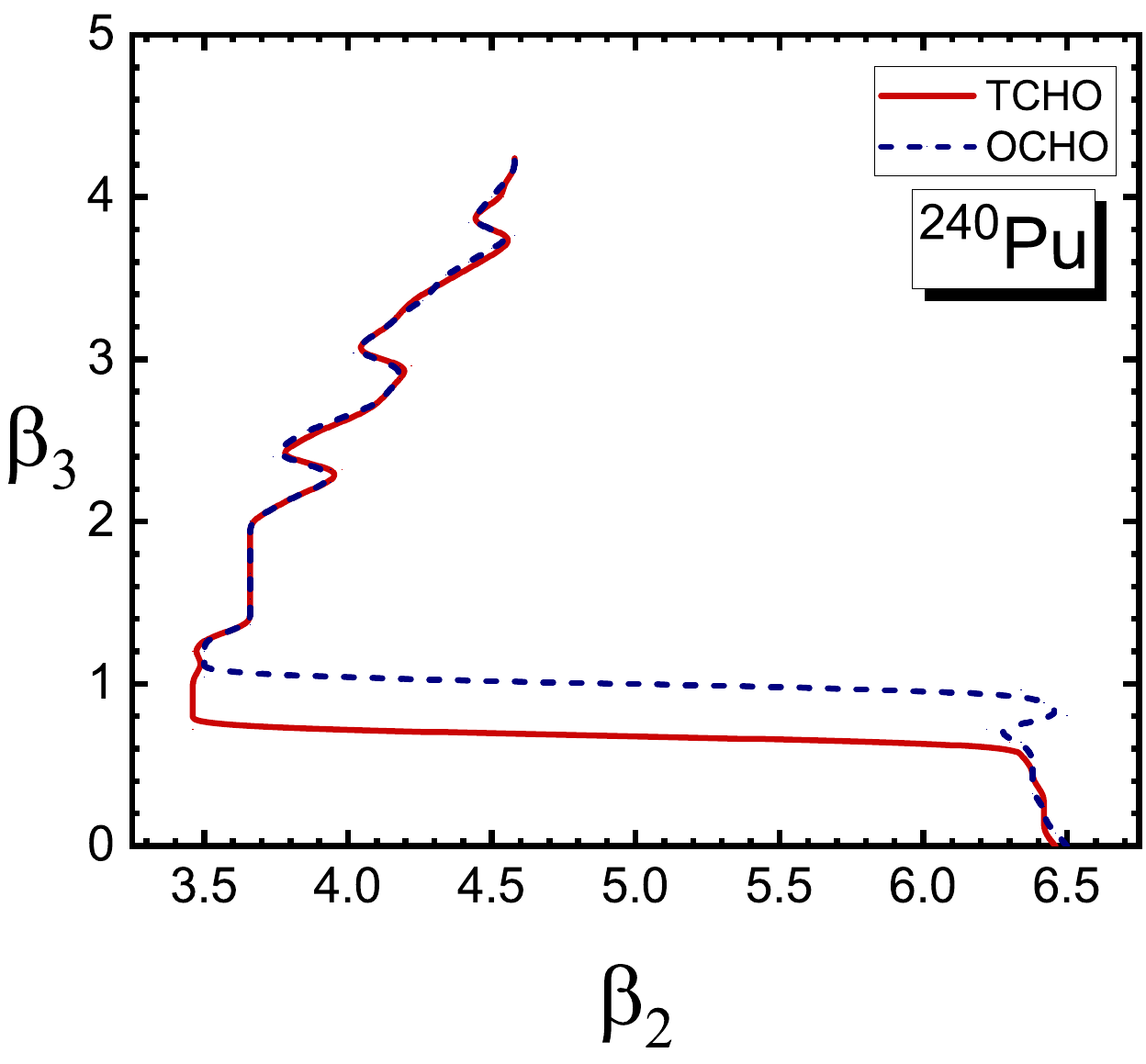}
  \caption{ (Color online) The scission lines of $^{226}$Th in the $\beta_2$-$\beta_3$ plane calculated using CDFT in TCHO and OCHO basis.}
  \label{scission-Th226}
\end{figure}

Fig. \ref{scission-Th226} displays the scission lines of $^{226}$Th in the $\beta_2$-$\beta_3$ plane calculated using CDFT in TCHO and OCHO basis. The patterns of them are very similar and there only a little difference at the ends of the symmetric and asymmetric fission valleys, i.e. $\beta_3\sim 0.0$ and $2.0$. In addition, we have also checked the mass tensor $B_{22}$ and $B_{33}$ calculated using Eq. (\ref{eq:BB}) and the relative root-mean-square discrepancy is within 0.03\%.

\begin{figure}[htbp]
  \centering
  \includegraphics[width=8cm]{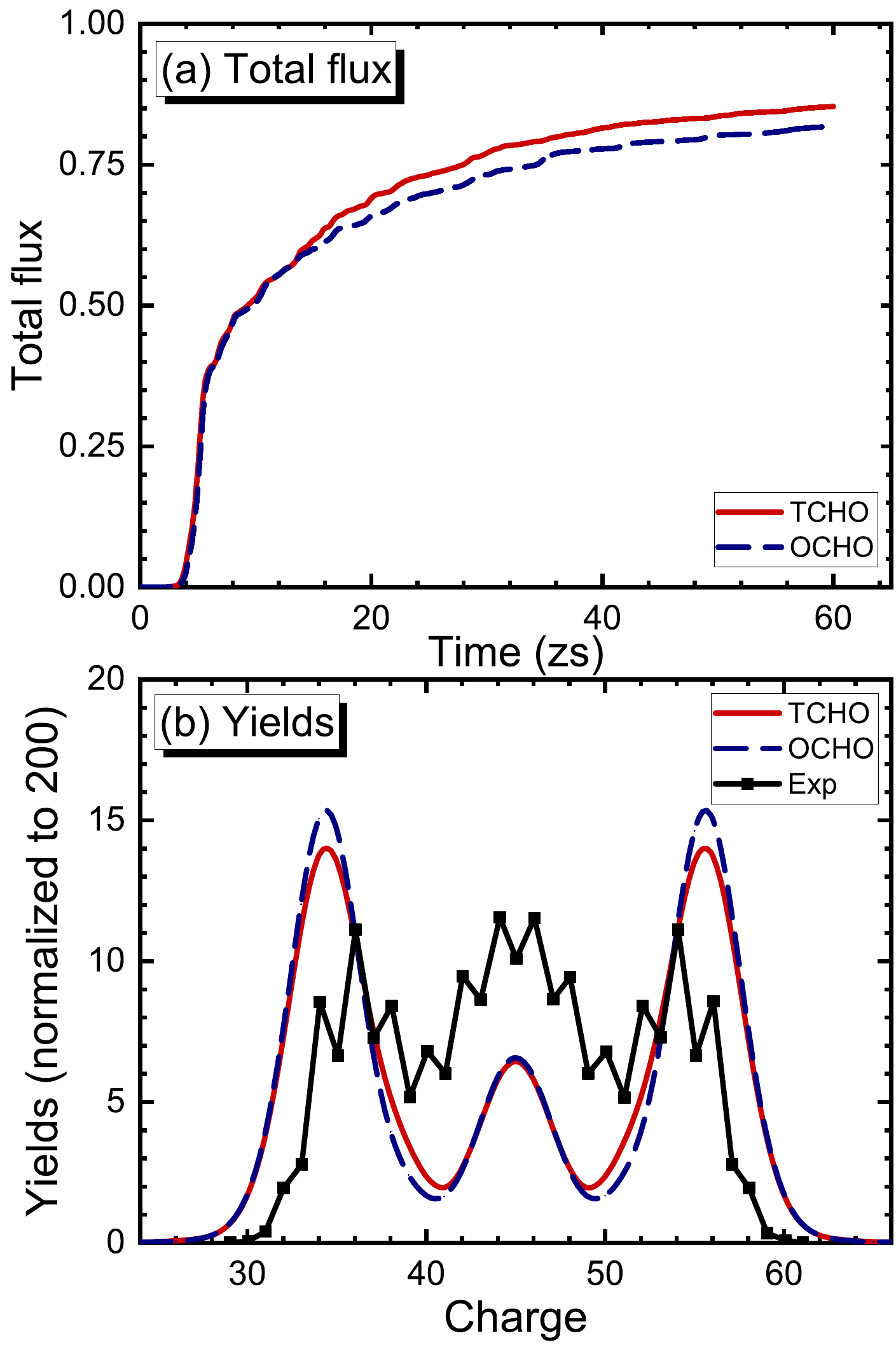}
  \caption{ (Color online) Total flux as a function of time (a) and  preneutron emission charge yields (b) for the photoinduced fission of $^{226}$Th calculated by TDGCM+GOA based on CDFT in OCHO and TCHO basis. The experimental charge yields are also shown for comparison \cite{Schmidt2000NPA}.}
  \label{td}
\end{figure}

Using the PESs, mass tensor, and scission configurations as inputs, we can simulate the dynamics for the photoinduced fission of $^{226}$Th in the framework of TDGCM+GOA. Following the procedure of Ref. \cite{TaoH2017PRC}, The initial state is prepared by boosting the collective ground state in the direction of $\beta_2$ with a target excitation energy about 11MeV \cite{Schmidt2000NPA}. Fig. \ref{td} displays the time evolution of total flux that passes through the scission line (panel a) and the preneutron emission charge yields (panel b) for $^{226}$Th calculated based on CDFT in OCHO and TCHO basis. Obviously, the total flux rises more rapidly for the calculation based on TCHO. This is easy to understand because the current is sensitive to the potential barrier, while a lower barrier is obtained by TCHO. The charge yields in Fig. \ref{td} (b) calculated based on both TCHO and OCHO  can reproduce the trend of the data, especially the coexistence of symmetric and asymmetric peaks in experimental data. Although the peak values are still away from the data, however, the TCHO calculation presents a trend to improve the description: The asymmetric fission peak value decreases from $15.07\%$ to $13.76\%$ and the dip at $Z\sim 40, 50$ increases from $1.66\%$ to $1.96\%$, which can be attributed to a reduction of the ridge between asymmetric and symmetric fission valleys (c.f. Fig. {PES}).

\subsection{Results for $^{240}$Pu}

\begin{figure}[htbp]
  \centering
  \includegraphics[width=9cm]{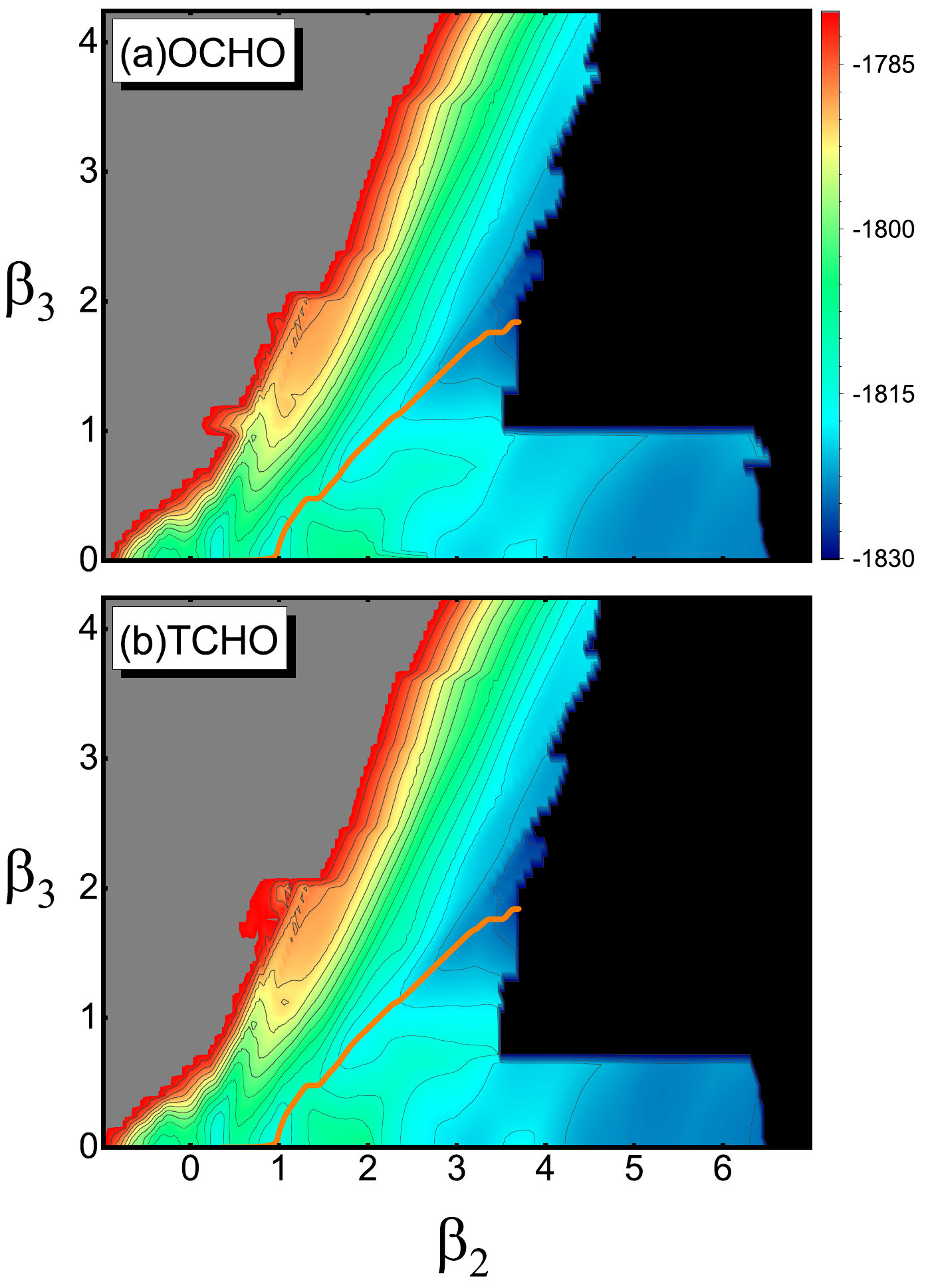}
  \caption{Same as Fig.\ref{PES} but for $^{240}$Pu}
  \label{PES_Pu240}
\end{figure}

\begin{figure}[htbp]
  \centering
  \includegraphics[width=9cm]{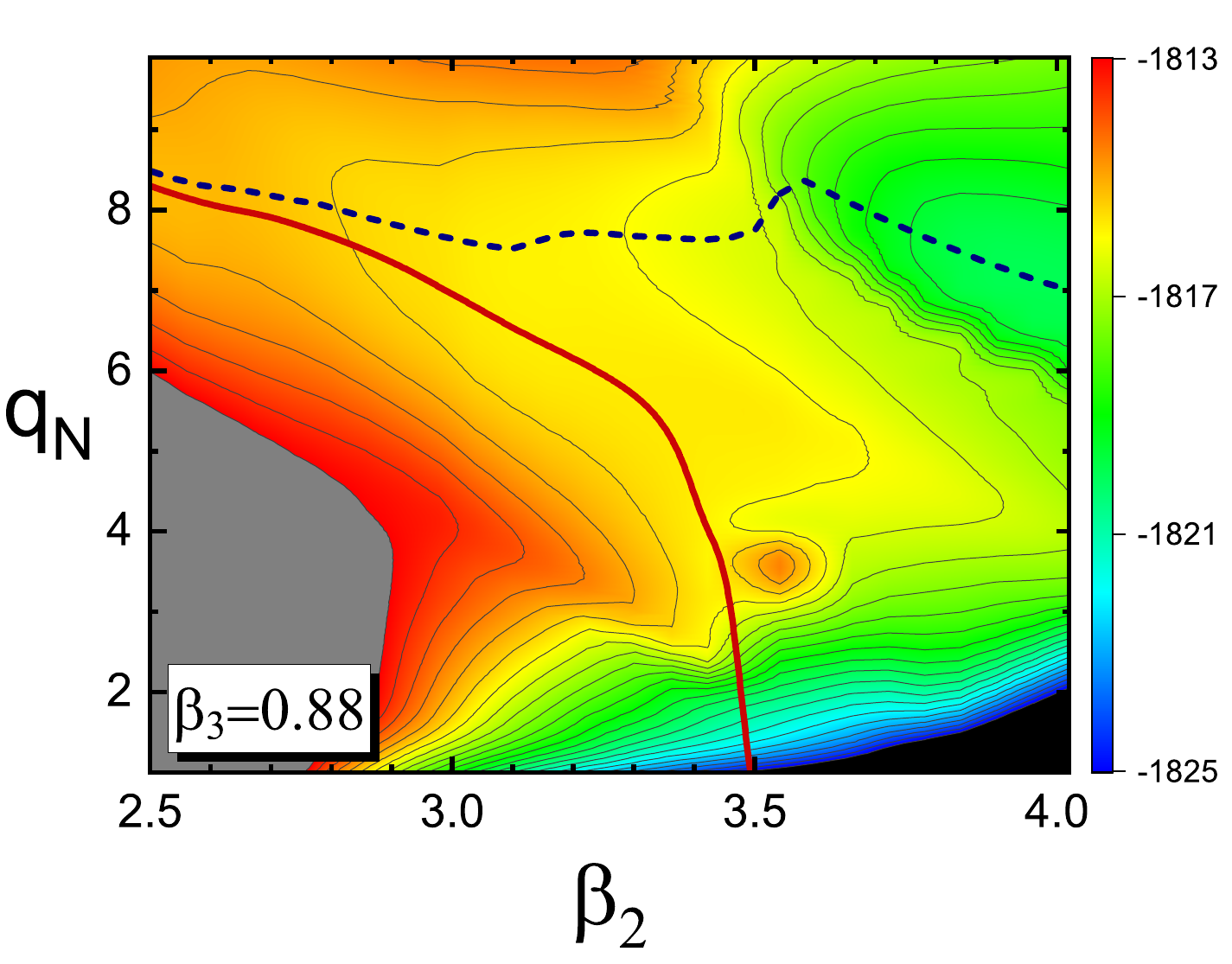}
  \caption{ (Color online) Potential energy surfaces of $^{240}$Pu in the $\beta_2-q_N$ plane for a fixed $\beta_3 = 0.88$ calculated by CDFT-TCHO with PC-PK1 functional. The solid curve denotes the fission path without constraint on $q_N$. For comparison, the corresponding fission path calculated by CDFT-OCHO is also shown as dashed curve.}
  \label{qn}
\end{figure}

Fig. \ref{PES_Pu240} displays the PESs of $^{240}\text{Pu}$ in the $\beta_2-\beta_3$ plane calculated using the CDFT in OCHO and TCHO basis. Along the static fission path, the barrier height of the first fission barrier is 6.79 MeV (OCHO) and 6.83 MeV (TCHO), while it is 3.62 MeV (OCHO) and 3.76 MeV (TCHO) for the outer barrier. It is noted that TCHO predicts higher fission barriers for $^{240}$Pu, different from the case in $^{226}$Th. For the region with $\beta_3\sim 0.88$, the calculated two PESs for $^{240}\text{Pu}$ are quite different: Scission happens much earlier in the TCHO calculation than that of OCHO. To understand the difference, we will analyze the PES in a higher dimension by introducing a constraint on the particle number in the neck $q_N$. Fig. \ref{qn} displays the PES in the $\beta_2$-$q_N$ plane with a fixed $\beta_3=0.88$ calculated by CDFT-TCHO. The solid curve denotes the fission path without constraint on $q_N$. For comparison, we also show the corresponding fission path calculated by CDFT-OCHO, denoted by the dashed curve. It is interesting to find that the nucleus goes to different fission valleys in two calculations due to the subtle difference.

\begin{figure}[htbp]
  \centering
  \includegraphics[width=8cm]{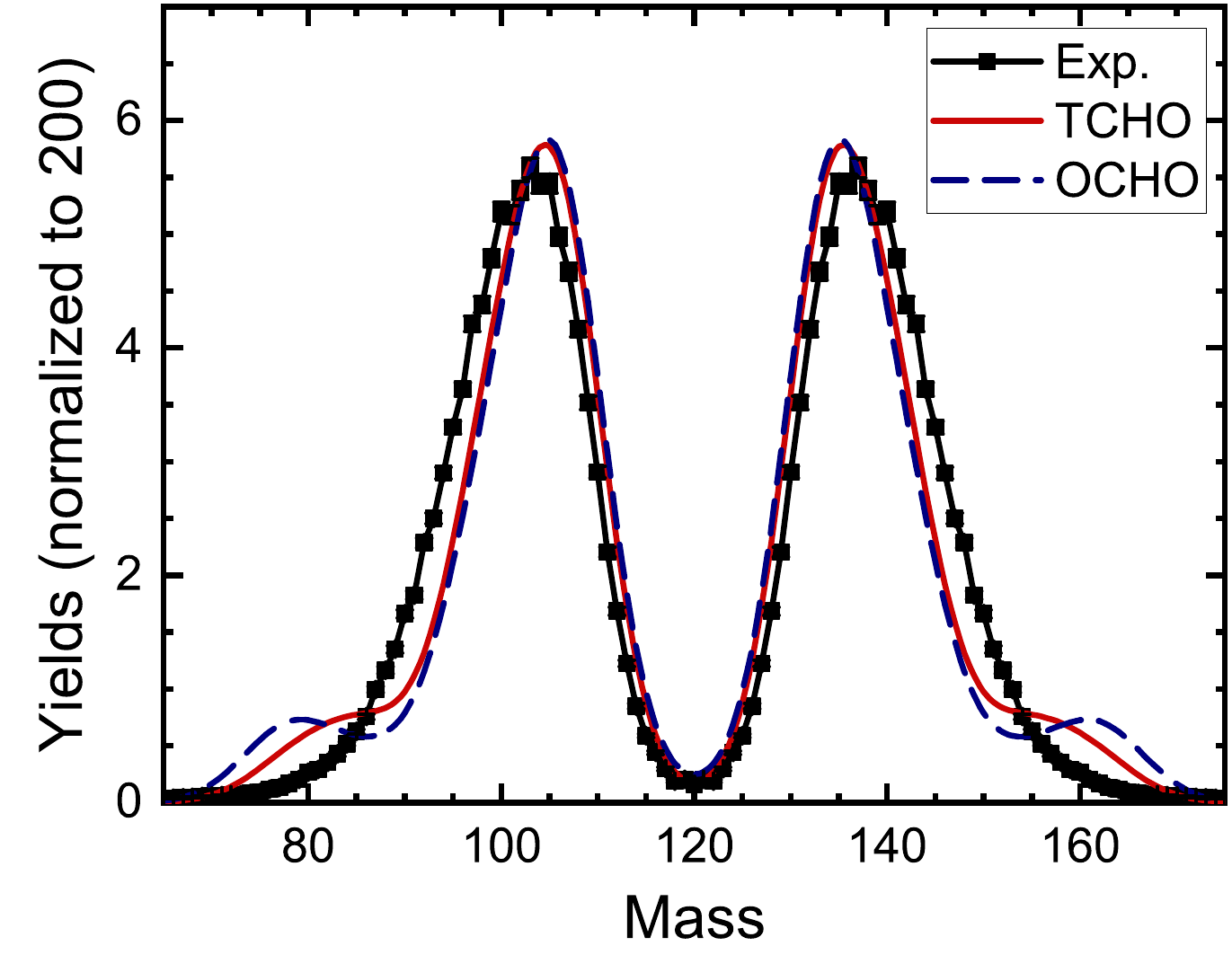}
  \caption{ (Color online) Mass distributions of the nascent fragments of $^{240}$Pu calculated by TDGCM+GOA based on OCHO and TCHO, in comparison with the experimental data \cite{C.Tsuchiya2000}.}
  \label{Yields_Pu240}
\end{figure}

The fission yields distribution calculated by TDGCM+GOA is shown in Fig. \ref{Yields_Pu240} and compared with the experimental data \cite{C.Tsuchiya2000}, while the initial state is prepared by simulating the initial state as a Gaussian superposition of collective eigenmodes in an extrapolated first potential well and the average energy lies at 1MeV higher than the inner fission barrier. The fission yield distribution presents double peaks, and both the peak values and peak positions are in good agreement with the experimental data. In the area with large mass asymmetry, two weak peaks are predicted in the calculation based on OCHO, which may be due to the too soft PES at the region with both large $\beta_2$ and $\beta_3$. Obviously, this has been modified in the calculation based on TCHO.

\section{Density constraint calculation for post-scission}\label{Density constrained DFT}

\begin{figure}[htbp]
  \centering
  \includegraphics[width=8cm]{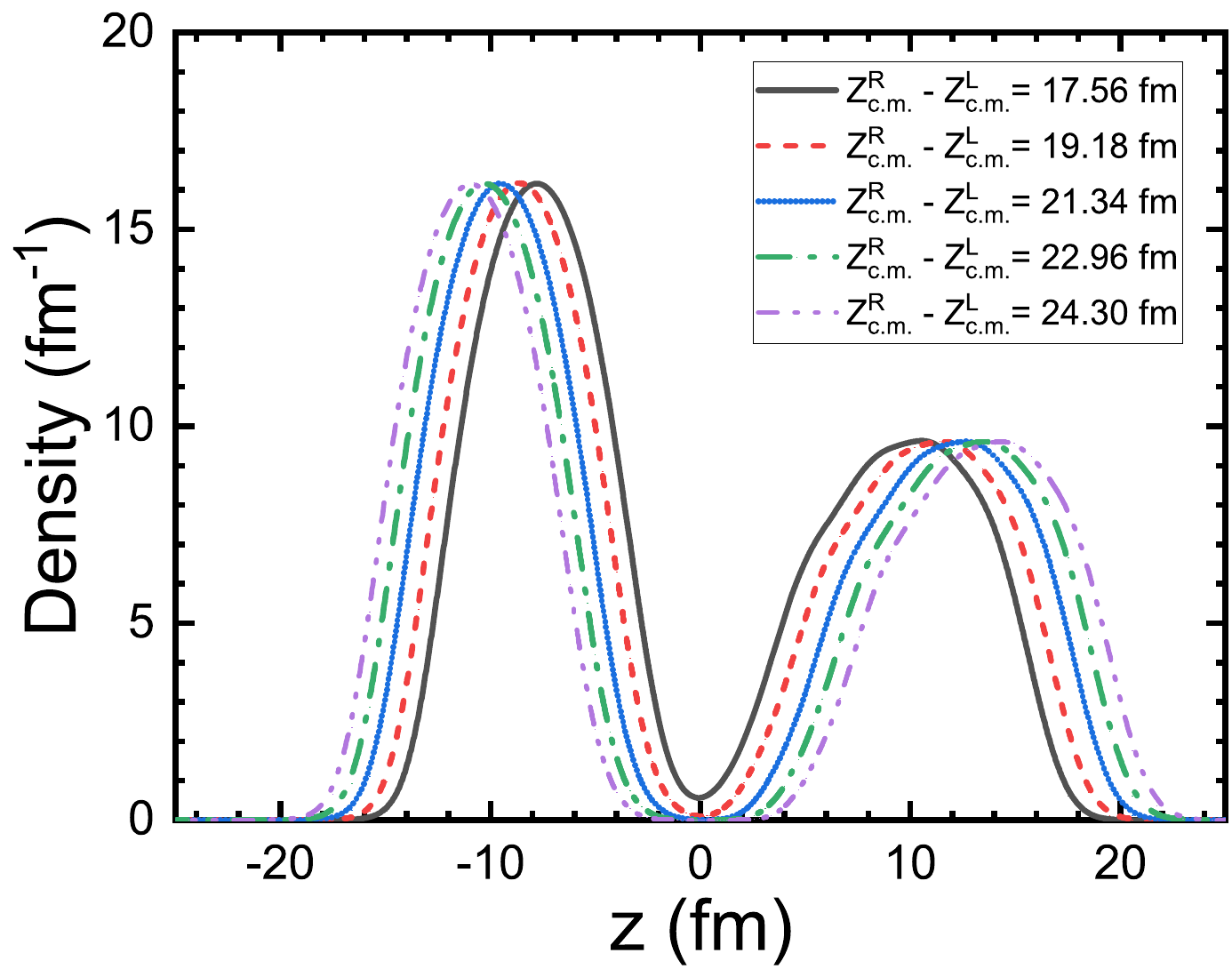}
  \caption{ (Color online) The density distributions of $^{240}$Pu along $z$ axis for the scission point and beyond. The black solid line denotes the density for scission point and the other lines from inside to outside indicate those for the configurations with the distance between the mass centers of the left and right nascent fragments gradually increases to 19.18, 21.34, 22.96, 24.30 fm.}
  \label{fragment}
\end{figure}

In a general fission process, starting from the excitation of the mother nucleus, scission usually occurs within 10$^{\text{-20}}$ seconds. Then, the fragments are far away from each other, and the fast neutron is emitted after about 10$^{\text{-17}}$ seconds. The huge difference between the two time scales allows us to assume that the configurations of the two nascent fragments will not change much in a short time after scission. Therefore, to simulate the splitting process of the nascent fragments beyond scission, we will introduce a density constraint in the new CDFT framework
\begin{gather}
  \langle E_{\rm tot} \rangle + C_\rho\left(\int \rho(r_\perp, z)-\rho_0(r_\perp, z)dr\right)^2
\end{gather}
where $C_\rho$ is the corresponding stiffness constant, $\rho$ and $\rho_0$ are the calculated and target densities of the nucleons, respectively. The target density can be chosen as the one of scission configuration but spliting the two fragments by a certain distance along $z$ direction with the center of mass fixed at $z=0$. Fig. \ref{fragment} displays the variation of the density distribution of $^{240}$Pu along $z$ axis from scission point to quite separated. Obviously, this method can simulate the fission procedure smoothly after scission and keep the properties of nascent fragments at scission.

\begin{figure}[htbp]
  \centering
  \includegraphics[width=8cm]{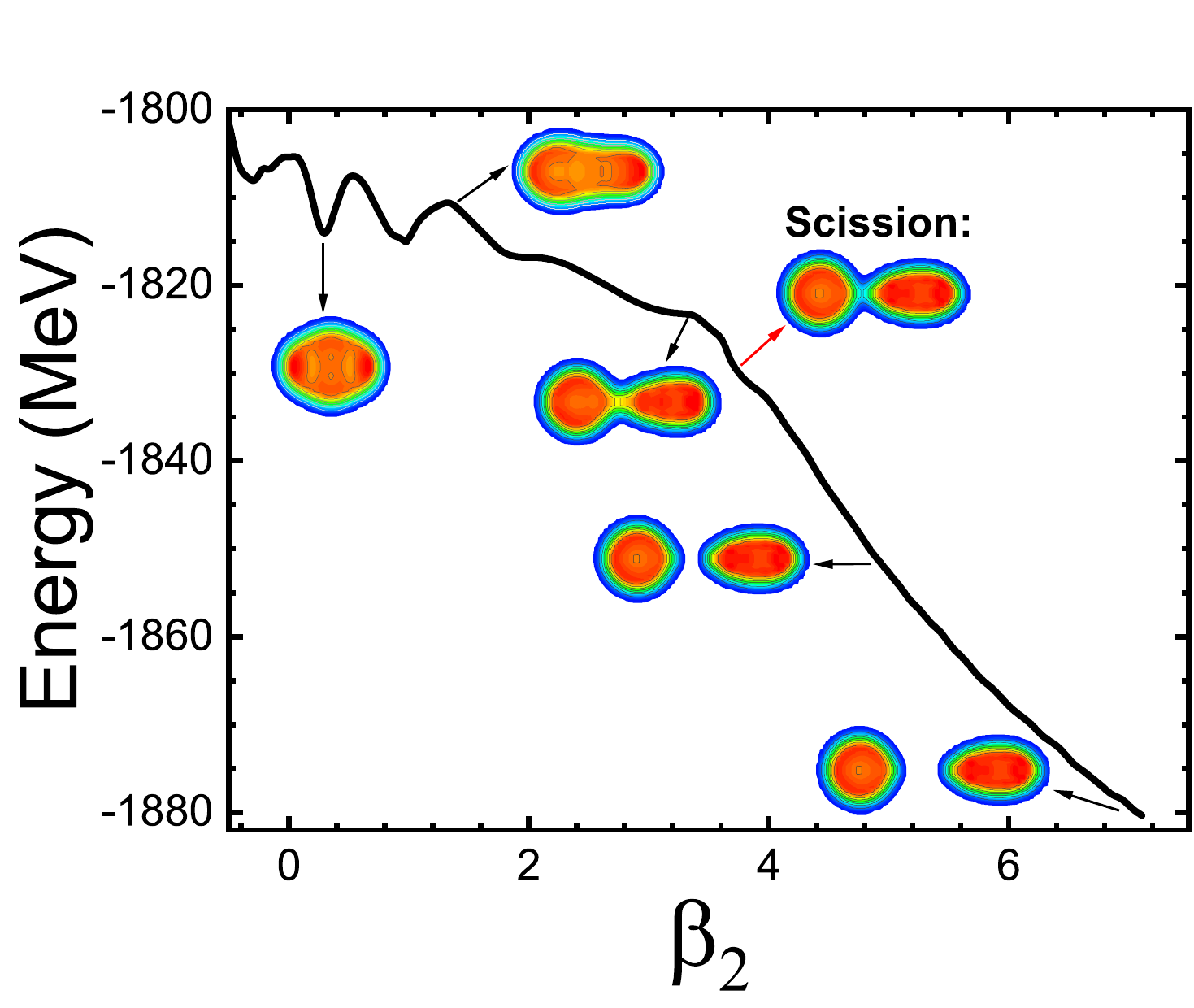}
  \caption{ (Color online) The potential energy curve of $^{240}$Pu along the optimal fission path including both prescission and postscission configurations. Contour plots of density distributions for some selected configurations are also shown.}
  \label{constrain_path}
\end{figure}

Figure \ref{constrain_path} displays the potential energy curve of $^{240}$Pu along the optimal fission path including both prescission and postscission configurations. Contour plots of density distributions for some selected configurations are also shown. Prescission configurations correspond to those denoted by the orange curve in Fig. \ref{PES_Pu240} (b), while the postscission configurations are obtained by density constraint calculation. Starting from the ground state with ellipsoidal deformation, the nucleus passes through two fission barriers, drops down rapidly to the scission point, and finally separates under Coulomb repulsion. This provides a full space to analyze the generation and evolution of the angular momentum of the fragments during the whole fission procedure, a work that is currently in progress.

\section{Summary and outlook\label{Summary}}

We have extended the point-coupling CDFT to be based on the TCHO basis and performed illustrative calculations for the PESs and induced fission dynamics of two typical examples: $^{226}$Th and $^{240}$Pu. A more reasonable PES is obtained in the new framework compared to that based on OCHO with the same basis space, especially for the outer fission barriers and large elongated configurations with an optimization of about $0.2\sim0.3$ MeV. Using the PESs, mass tensor, and scission configurations as inputs, we have also simulated the dynamics for the induced fission of $^{226}$Th and $^{240}$Pu in the framework of TDGCM+GOA. The calculations based on TCHO basis presents a trend to improve the description for fission yields. Finally, we also introduced a density constraint into the new framework to simulate the postscission procedure by separating the two frozen fragments from scission point. This provides a full space to analyze the generation and evolution of the angular momentum of the fragments during the whole fission procedure, a work that is currently in progress.

The new developed CDFT-TCHO optimizes the elongated configurations, improves the calculation efficiency, and provides a basis for large-scale multi-dimensional constraint calculation. Very recently, we have performed a fully three-dimensional (3D) calculation to generate the 3D PES for the fission of compound nucleus $^{236}$U using CDFT-TCHO with constraints on the axial quadrupole and octupole deformations $(\beta_2, \beta_3)$ as well as the nucleon number in the neck $q_N$ \cite{Zhou_2023}. By considering the additonal degree of freedom $q_N$, the PES broadens up to form a wide ``estuary'' in the $(\beta_2, q_N)$ subspace for $q_N<6$: the energy surface is very shallow across a large range of quadrupole deformations. This leads to a fluctuation for the estimated total kinetic energies by several to ten MeV and for the fragment masses by several to about ten nucleons. Of course, this is just a simple estimation for the fluctuation of the fission observables. More precise analysis should be done by performing 3D TDGCM+GOA calculation based on the 3D PES.

\begin{acknowledgments}
This work was partly supported by the National Natural Science Foundation of China (Grants No. 11875225, No. 11790325,  No. 11790320),  the Fundamental Research Funds for the Central Universities, and the Fok Ying-Tong Education Foundation.
\end{acknowledgments}

\appendix
\section{\label{appendixA}Determination of the parameters in TCHO basis}

In Sec. \ref{subsec:TCHO}, we have introduced the TCHO basis and the expansion of Dirac equation in detail. Here, we will briefly introduce how to determine the two parameters $z_1$ and $b_1$ in TCHO basis.  Fig. 12 displays the total binding energies for some selected configures of $^{226}$Th optimal fission path calculated by CDFT-TCHO using different basis parameters  $z_1$ and $b_1$. All energies are normalized with respect to the binding energy of the corresponding global minimum. Obviously, we can observe some regions with lowest binding energies in the $z_1-b_1$ plane, and these regions move to large $z_1$ but keep around $b_1\sim 3.3$ fm as the deformation increases. For the convenience of practical application, we have comprehensively considered the symmetric fission path and the optimal fission path, and summarized the following formulas for the two parameters
\begin{gather}
  z_1=\left\{ \begin{split}
    0~~~~~~~~~~~~~~~~~~~~~~~~~~~~~~~~~~~~~~~~~~~~~~~~~~&\beta_2<1  \\
    3.05\sqrt{1.68{{\beta }_{2}}-1.40}-1.60~~~~~~~~~~~~~~~~~&\beta_2>1    \\
 \end{split} \right. \label{basis_equ1} \\
  b_1=\left\{ \begin{split}
    b_z~~~~~~~~~~~~~~~~~~~~~~~~~~~~~~~~~~~~~~~~~~~~&\beta_2<0    \\
    (b_z-3.3)\beta_2^2+(6.6-2b_z)\beta_2+b_z~~~~~~~&0<\beta_2<1  \\
    3.3~~~~~~~~~~~~~~~~~~~~~~~~~~~~~~~~~~~~~~~~~~~~&\beta_2>1    \\
 \end{split} \right. \label{basis_equ2}
\end{gather}
where $b_z$ is the corresponding characteristic length related to the frequency of the spherical harmonic oscillator $\hbar\omega=41 A^{-1/3}$. For small deformations, $z_1$ is set as 0, namely going back to OCHO, and $b_1$ changes gradually from 3.3 fm to $b_z$. The efficiency of these formulas has also been approved in the calculations for a number of heavy nuclei.

\begin{figure}[htbp]
  \centering
  \includegraphics[width=8.5cm]{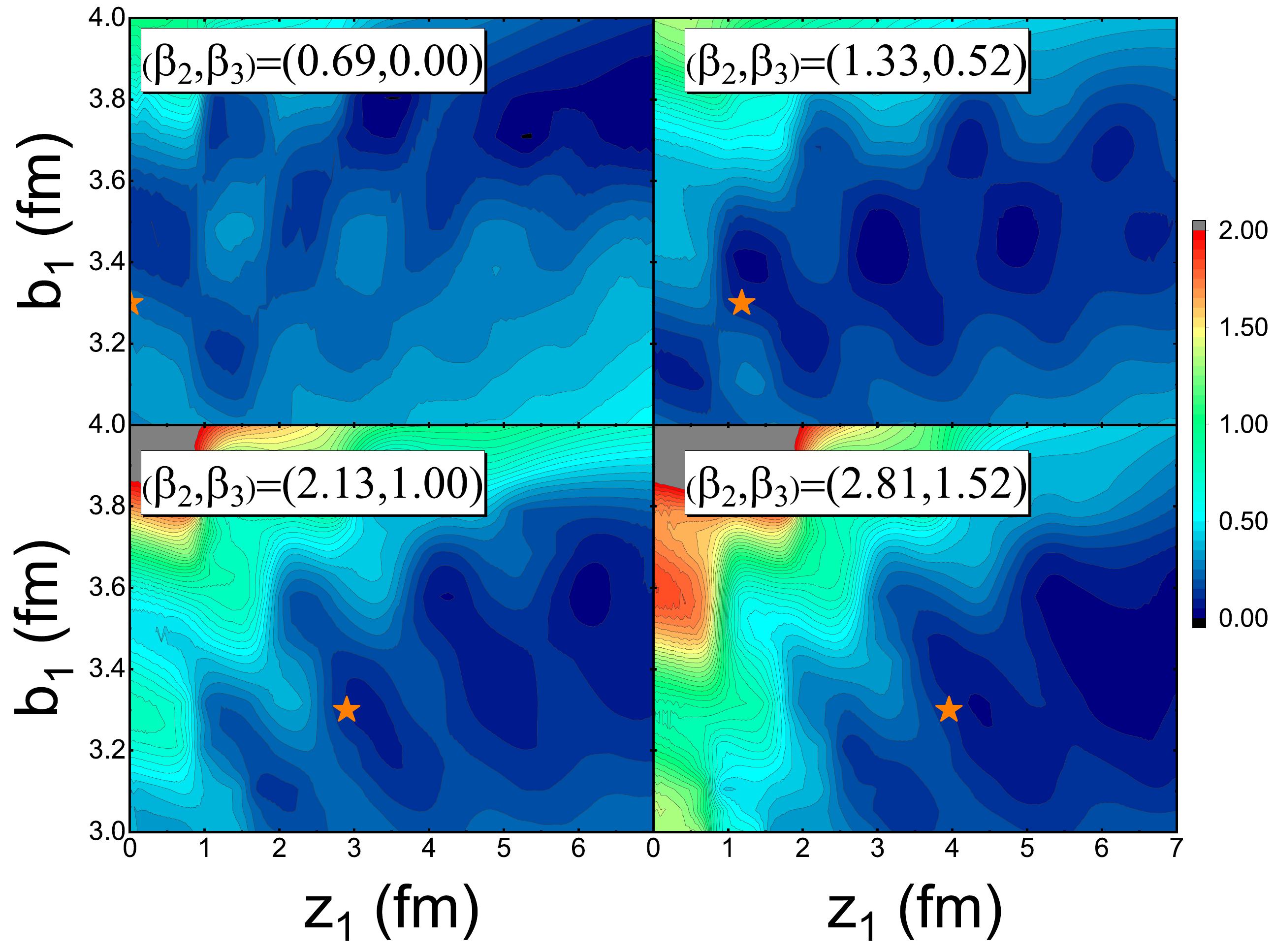}
  \caption{The total binding energies for some selected configures of $^{226}$Th optimal fission path calculated by CDFT-TCHO using different basis parameters  $z_1$ and $b_1$. All energies are normalized with respect to the binding energy of the corresponding global minimum. The energy difference between adjacent contour lines is 0.05 MeV. The stars denote the basis parameters determined by Eqs. (\ref{basis_equ1}, \ref{basis_equ2}).}
  \label{para}
\end{figure}

\newpage

\nocite{1}
\bibliography{TCHO}

\end{document}